\documentclass[prl,aps,twocolumn]{revtex4}
\usepackage{graphicx}
\usepackage{subfigure}
\usepackage{bm}
\usepackage{amsmath}
\begin{document} 
\title{Dynamical tunneling in molecules: role
of the classical resonances and chaos}
\author{Srihari Keshavamurthy}
\affiliation{Department of Chemistry, Indian Institute
of Technology, Kanpur, India 208 016}
\date{\today}
\begin{abstract}
In this letter we study dynamical tunneling in 
highly excited symmetric molecules. 
The role
of classical phase space structures like resonances and chaos on the
tunneling
splittings are illustrated using the water molecule as an example. 
It is argued that the enhancements in the splittings due to
resonances (near-integrable phase space) and due to chaos (mixed phase space)
are best understood
away from the fluctuations associated with avoided crossings. 
In particular we provide an essential difference between the two mechanisms
in terms of high order perturbation theory. 
The analysis, apart from testing the validity of a perturbative approach,
suggests such systems as prime candidates for studying dynamical tunneling.
\end{abstract}
\maketitle

The concept of dynamical tunneling emerged more than
two decades ago in the context of
studying near degenerate vibrational states of symmetric molecules.
As an example in the water molecule 
the symmetry of the OH bonds
implies that a state $(n_{1} = 3,n_{2} = 0)$
which has three quanta of excitation in one of
the OH bond ($n_{1} = 3$)
and zero quanta of excitation in the other equivalent OH
bond ($n_{2} = 0$) is degenerate with 
the state $(0,3)$.
The two states are not coupled through any classical process 
{\it i.e.,} a classical trajectory started with initial conditions
corresponding to the state $(3,0)$ will remain localized 
indefinitely without ever evolving to the part of the phase space
corresponding to the state $(0,3)$.
However, Lawton and Child\cite{lawch}
realized that a generalized form of tunneling mixes the two degenerate states
giving rise to a characteristic splitting.
In an influential work Davis and Heller\cite{davhel} refined and
generalized the earlier observations and introduced
the term dynamical tunneling to distinguish it from the usual
tunneling through potential barriers in coordinate space.
One of the important outcomes of the study was the suggestion that
the underlying phase space of the system
is the proper setting to understand dynamical tunneling.

In an apparently unrelated developement researchers studying
the problem of intramolecular energy 
redistribution (IVR) discovered purely quantum
energy flow between modes which would be otherwise uncoupled.
A detailed analysis by Hutchinson, Sibert and Hynes\cite{husihy}
established that the mechanism for such clasically forbidden 
energy flow between degenerate vibrational modes arose from indirect
state-to-state explorations involving a sequence of intermediate states.
Important insights were provided by 
Stuchebrukhov and Marcus\cite{stumar} 
who established that one could view dynamical
tunneling as a high order perturbative process involving a sequence
of off-resonance virtual states (``vibrational superexchange"). 
Interestingly it was also shown that the perturbative approach to 
the tunneling splitting was related in a simple way to the usual
semiclassical JWKB solution.

Despite this important prior work the explicit demonstration of the
role of various classical phase space structures in dynamical tunneling
has only recently been established.
In particular the work of Bohigas, Tomsovic and Ullmo\cite{cat}
elucidating the role of chaos on dynamical tunneling 
has led to a resurgence of interest in the field.
A key observation that emerged from 
numerous studies\cite{cats} of chaos-assisted
tunneling (CAT) is that such a process necessarily requires
atleast a three level mechanism. One of the hallmarks of CAT has to do with
erratic fluctuations of the tunnel splittings with variations in energy
or system parameters. The fluctuations were explained on the basis of
a tunnel doublet, associated with
regular regions in the phase space, invoved in an avoided 
crossing with a third state
inhabiting the chaotic region of the phase space.

However, very recently  
classical nonlinear resonances
in the near-integrable phase space regimes
have also been implicated
to play a dominant role in dynamical tunneling\cite{rat1,rat2}.
This has been dubbed, in analogy with CAT, as 
resonance-assisted tunneling (RAT)
and studies revealed that tunnel splitting fluctuations,
possibily more intense than in CAT,
occur in near-integrable systems as well due to crossing
of the regular tunnel doublets with a regular third state.
The sheer richness of the phase space perspective in dynamical tunneling is
further exemplified by the work of Frischat and Doron\cite{frido}.
At present the consensus regarding dynamical tunneling
seems to be that CAT and RAT are manifestations of the more
general phenomenon of transport-assisted
tunneling\cite{frido}.

The preceeding discussion makes it clear that splitting fluctuations 
can occur in general nonintegrable systems
and tunneling can be ascribed to chaos or resonances
depending
on the phase space nature of the third 'intruder' state.
It therefore seems natural to focus attention
away from the
avoided crossings in order to gain insights into the process\cite{rat2}. 
The present work is concerned with the study of dynamical tunneling
in molecules from such a perspective.
An essential difference from the work of Brodier, Schlagheck and
Ullmo\cite{rat2} is that the focus in this work is entirely
on a perturbative evaluation of the splittings to quantitative
accuracies. 
The splittings $\Delta$ 
associated with the doublet states $|i\rangle$
and symmetry related $|f\rangle$
are calculated by high order nondegenerate perturbation theory
involving the unperturbed states $\{\alpha_{r}\}$ as
\begin{equation}
\frac{\Delta}{2} = 
\sum_{r \neq i,f}^{} 
\frac{V^{(1)}_{i1}V^{(2)}_{12} \ldots V^{(n)}_{nf}}
{\Delta E_{i1}\Delta E_{i2}\ldots \Delta E_{nf}},
\end{equation}
where $\widehat{V}^{(r)}$ are the local
perturbations that connect the intermediate states
through  
$V^{(r)}_{qr} = \langle \alpha_{q}|\widehat{V}^{(r)}| \alpha_{r}\rangle$
with $\Delta E_{ir}= 
(E_{i}-E_{r})$.
The sum over the intermediate states can be viewed as a sum over paths
in the state space and the number of intermediate states is related to
the length of a path. 
The perturbative treatment is valid as long as $V^{(r)}_{qr}/2 <<
\Delta E_{ir}$ and hence the method cannot be used to calculate
splittings for states involved in avoided crossings.
In principle, an infinite number of high order perturbative
chains (paths in state space) exist which connect the two degenerate
states. The hope has been that perhaps a few such paths
would dominate the dynamical tunneling in the integrable and near-integrable
cases.
In this work we show that even in near-integrable
cases there is no single dominant perturbative path. 
However, it is possible to identify a certain family of
perturbative chains (``minimal paths" in state space) that are
sufficient to provide quantitative results. 
Parameteric variations
tune the phase space 
from the near-integrable limit to the mixed regime 
and it is observed
that the minimal paths are not sufficient anymore to reproduce the splittings. 

\begin{figure}
\includegraphics*[width=2in,height=2in]{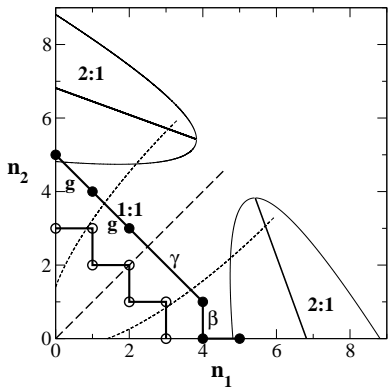}
\caption{State space and the location of the various resonances for
water with $P=8$. The 2:2 zone has been supressed for clarity.
Two minimal perturbative paths $\Gamma_{600}$ (open circles)
and $\Gamma_{221}$ (filled circles)
in state space are shown as
examples.}
\label{fig 1}
\end{figure}

The purpose of this letter is to illustrate the observations above
and accentuate the various mechanisms of dynamical tunneling.
The system of choice is the 
water molecule which can be described by an effective spectroscopic 
Hamiltonian\cite{bagg}
\begin{equation}
\widehat{H}_{\rm eff} = \widehat{H}_{0} + 
 g \widehat{V}^{(12)}_{1:1} + 
\gamma \widehat{V}^{(12)}_{2:2} + \beta(\widehat{V}^{(1b)}_{2:1} +
\widehat{V}^{(2b)}_{2:1}),
\end{equation}
where $\widehat{H}_{0}$ is 
diagonal in the number basis $(n_{1},n_{2},n_{b})$
and $g,\gamma,\beta$ 
represent the strengths of the various perturbations.
The various parameters of $\widehat{H}_{\rm eff}$ are given
in the work of Baggott\cite{bagg}.
The zeroth order quantum
numbers $(n_{1},n_{2},n_{b})$ represent the excitation quanta in the
two, equivalent OH-stretches and the bend mode respectively. The 
perturbations $\widehat{V}$ are off-diagonal in the 
number basis and have the form:
\begin{equation}
\widehat{V}^{(ij)}_{m:n} =  
\left[(\hat{a}_{i})^{n}(\hat{a}_{j}^{\dagger})^{m} + {\rm h.c.}\right]
\end{equation}
where $\hat{a}_{i}$ and $\hat{a}_{i}^{\dagger}$ are the
annhilation and creation operators for the mode $i$.
The resonant perturbations $\widehat{V}^{(ij)}_{m:n}$
are responsible for the exchange of
quanta (energy) between the modes $i$ and $j$.
Classical-quantum correspondence studies\cite{ksjcp} have established
that the classical limit Hamiltonian corresponding to 
$\widehat{H}_{\rm eff}$ is nonlinear and multiresonant. 
The location of the various resonance zones\cite{ksjcp}
in state space are shown in
Fig.~(\ref{fig 1}). Note that the previous studies\cite{lawch,stumar} 
on water correspond
to integrable phase space regimes.

Clearly $\widehat{H}_{\rm eff}$ is 
symmetric under $n_{1} \leftrightarrow n_{2}$ 
and 
$P \equiv n_{1}+n_{2}+n_{b}/2$ is a conserved quantity. 
Consequently $\widehat{H}_{\rm eff}$ 
for a given $P$ is of the order $(P+1)(P+2)/2$ making the
numerical investigations rather straightforward.
In addition $\widehat{H}_{\rm eff}$ exhibits a wide range of
behaviour from integrable to near-integrable to mixed classical
dynamics. Due to the exact $C_{2v}$
symmetry tunneling doublets
exist over the
entire range of the classical dynamics. 
In this work six tunneling doublets with $P=8$, denoted by
$|m,\pm m,n_{b}\rangle$ with $m=n_{1}+n_{2}$ 
are studied. 
The labels used for the states are appropriate in the absence of
the 2:1 resonances. We will continue to use them for convenience
keeping in mind that
the actual assignments of these states are different\cite{ksjcp}.
The energies of the states approximately span the range
$[24256,25574]$ cm$^{-1}$ above the ground state.
For comparison the various resonant strengths 
have the typical values $\beta \approx 27$ cm$^{-1}$,
$g \approx -50$ cm$^{-1}$ and $\gamma \approx -1.0$ cm$^{-1}$.

\begin{figure}
\includegraphics*[width=2.5in,height=2.5in]{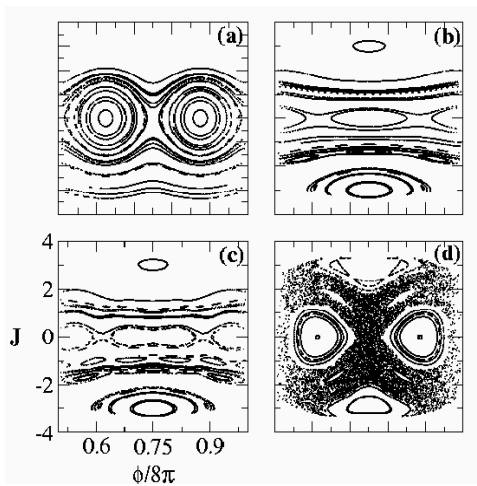}
\caption{Poincar\'{e} surface of sections at energy corresponding
to the doublet $|6,\pm 6,4\rangle$. The action variable
$J=(I_{1}-I_{2})/2$ and the conjugate angle variable $\phi = \theta_{1}-
\theta_{2}$. (a) Integrable 1:1+2:2 (b) Near-integrable 2:1
(c) Near-integrable 2:1+2:2 and (d) full system.
The 2:1 islands are centered at
$\phi = 6 \pi$ and $J \approx \pm 3$ while the 1:1 islands appear
around $J = 0$. The surface of section is symmetric with respect
to reflection about the $J=0$ axis.}
\label{fig 2}
\end{figure}

\begin{figure}
\includegraphics*[width=2.5in,height=2.5in]{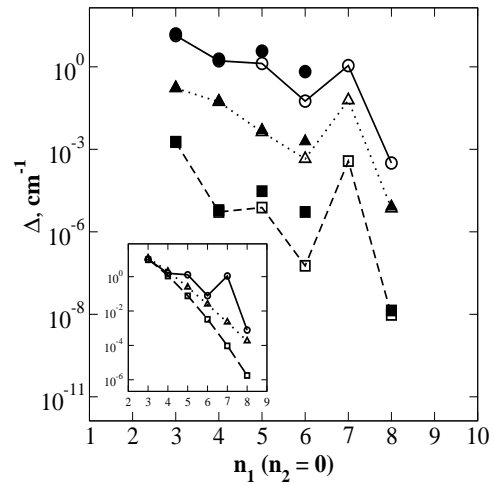}
\caption{Dynamical tunneling splittings for
the nonintegrable cases. Squares (only 2:1),
triangles (2:1+2:2), and 
the full system (circles).
The perturbative results (converged) using
minimal paths are shown by
filled symbols. Mean level spacings $\approx$ 100 cm$^{-1}$.
(Inset) Splittings for the integrable
cases of the 1:1 (squares) and 1:1+2:2 
(triangles). The high order perturbation
results are shown as dashed and dotted lines
respectively. The full case is also shown for comparison.}
\label{fig 3}
\end{figure}

Reperesentative phase space sections are shown in
Fig.~(\ref{fig 2}) for the highest energy state among the
doublets considered.
In Fig.~(\ref{fig 3}) the doublet splittings, $\Delta_{m}$,
are shown for the various states $|m,\pm m,n_{b}\rangle$.
Note that the states in Fig.~(\ref{fig 3}) are not energy ordered
and in particular the state $|8,\pm8,0\rangle$ is lower in energy
than the state $|5,\pm 5,6\rangle$.
The splittings for the integrable cases (only 1:1 and 1:1+2:2)
are shown in the inset to Fig.~(\ref{fig 3}).
As expected\cite{stumar} the splittings show 
a monotonic decrease with increasing amount
of stretch excitation and no fluctuations are observed.
Even within the integrable cases it is relevant to note that
the addition of the weak 2:2 results in a significant increase in
$\Delta$.
High order perturbation calculation\cite{stumar}
essentially reproduces the
splittings to quantitative accuracy as is apparent from
the inset in Fig.~(\ref{fig 3}).
Inclusion of the 2:1 stretch-bend resonances renders the system
nonintegrable and clear deviations from the integrable cases are seen.
In the main part of Fig.~(\ref{fig 3}) the splittings are
shown as one proceeds from near-integrable to the full, mixed scenario.
The nature of the underlying phase space clearly reflects the
transition from near-integrable to the mixed 
regime (cf. Fig.~(\ref{fig 2}b-d)).
The main difference from the integrable case is the
nonmonotonic behaviour of $\Delta_{m}$. The intense fluctuations for the
state $|7,\pm 7,2\rangle$ are associated with its proximity to an
avoided crossing. The key observation here is that the fluctuation
due to avoided crossing is a robust feature and persists in the full system.
Indeed it has been confirmed that in the 2:1 only case there
is a strong doublet-doublet crossing which transforms into a singlet-doublet
crossing in the full system and for 
the deuterated analog D$_{2}$O a similar calculation
shows absence of fluctuations in the
full system despite the presence of significant chaos\cite{tbp}.
Thus it is crucial to
shift the focus away from the fluctuations to uncover
the effect of chaos and resonances on dynamical tunneling
in the system.

In Fig.~(\ref{fig 3}) the phenomenon of RAT is
evident in going from the 2:1 only system to the 2:1+2:2 system.
The enhancements of the splittings are even more dramatic
considering the fact that states with odd $m$
do not split at all if only the 2:2 resonance is present.
Similarly the enhancement of the splittings on going from the 
2:1+2:2 to the full system can be ascribed to CAT.
Note that the enhancement due to CAT are typically an order of magnitude
smaller than those due to RAT as has been earlier observed in
the annular billiard studies\cite{frido}.
A crucial difference between CAT and RAT arises from the perturbative
viewpoint. 
In order to highlight this difference we begin by noting that
the structure of the resonances imply, in terms of
order, $\beta^{4} \sim g^{2} \sim \gamma$.
Hence a consistent perturbative calculation of the
splitting for the state $|m,\pm m,n_{b}\rangle$ is obtained as:
\begin{equation}
\Delta_{m} = \sum_{a,b,c}^{} \beta^{a} g^{b} \gamma^{c} \sum_{n}^{}
\Delta_{m}(\Gamma_{abc}^{(n)})
\end{equation}
where $n$ indexes all possible paths $\Gamma_{abc}$ for a particular choice
of $a,b,c$ satisfying the constraint $a+2b+4c=2m$.
It is clear that all possible paths have not been included in the above
calculation of $\Delta_{m}$. The paths included, called
minimal paths, are all of the effective
order $2m$. 
Examples of the minimal paths $\Gamma_{600}$ and $\Gamma_{221}$ 
relevant for calculating $\Delta_{3}$ and $\Delta_{5}$ respectively are
shown in Fig.~(\ref{fig 1}).
One of the observations from this work is that for near-integrable
systems a particular family of the minimal paths is dominant. 
In contrast, 
more than one family 
and higher order paths are required in the presence of significant chaos.
As an example consider the case of $\Delta_{5}$. 
This state is involved in a weak avoided crossing and hence provides
a stringent test for the nondegenerate perturbation theory.
In the near-integrable cases the family of
paths $\Gamma_{10 00}$, $\Gamma_{202}$, and $\Gamma_{601}$ 
are sufficient.
Perturbative calculations yield the splittings
$\Delta_{5}(\Gamma_{10 00}) = 6.1 \times 10^{-5}$ cm$^{-1}$ and 
$\Delta_{5}(\Gamma_{202,601}) = 7.2 \times 10^{-3}$ cm$^{-1}$ as
compared to the numerically exact values of 
$1.7 \times 10^{-5}$ and $6.2 \times 10^{-3}$ cm$^{-1}$ respectively.
As mentioned earlier the 2:1 only system is less accurate due to
the avoided crossing.
In the integrable cases the 
perturbative results $\Delta_{5}(\Gamma_{050}) \approx 7.0 \times
10^{-2}$ and $\Delta_{5}(\Gamma_{050,031,012})
\approx 0.23$ cm$^{-1}$ are in good agreement with the exact
values of $6.7 \times 10^{-2}$ and $0.23$ cm$^{-1}$ respectively.
However, for the full system neither the near-integrable 
nor the integrable families are sufficient to reproduce the
splitting ($\approx 1.6$ cm$^{-1}$). 
In fact the families $\Gamma_{430},\Gamma_{240},\Gamma_{221}$
play an important role and yield a splitting of about $3.4$ cm$^{-1}$.
For mixed systems, in general,
a large number of families in the minimal set
as well as higher order paths have to be included for a quantitatively
accurate calculation
of $\Delta_{m}$. 
Thus, for instance,
a minimal path calculation of $\Delta_{6}$ yields a splitting
which is about an order of magnitude large. 
The correlation of the preceeding observation with increasing amount of
chaos in the phase space is an interesting possibility\cite{tbp}.
As an example we mention that tuning the parameter $\gamma$ from
about $1 \rightarrow 4$ cm$^{-1}$ turns the near-integrable
$2:1+2:2$ system into a mixed system. The splitting $\Delta_{5}$ is seen
to increase by about a factor of $22$ whereas the minimal perturbative
calculation predicts an accelaration by a factor of $12$.
Clearly one anticipates higher order paths to play a significant role
in mixed systems.

In conclusion we have shown that molecular systems with symmetry
are prime candidates to study both chaos and resonance assisted dynamical
tunneling\cite{orti}. 
High resolution spectroscopic studies of rovibrationally excited
states are typically 
analysed in terms of such effective Hamiltonians\cite{hrs}.
The splitting patterns are crucial for understanding the
extent of IVR in the molecule and 
hence the importance of dynamical
tunneling\cite{ejhsar}. 
This work is the first one to perform a detailed analysis
of the possible mechanisms of dynamical tunneling in molecules
with the hope that both CAT and RAT can be experimentally observed
in the high resolution spectra. 
We have argued that a clear distinction
between CAT and RAT can be established by focusing attention away from
the avoided crossings. In addition high order perturbation theory, when
valid, is quite successful in reproducing the tunnel splittings in the
case of integrable and near-integrable systems. A single family
of perturbative paths is sufficient but it is important to point out
that among the many individual paths possible there is no 
single dominant path. 
For instance in the calculation of $\Delta_{5}$ the family
$\Gamma_{1000}$ has $242$ paths which come with both positive and negative
contributions and one needs to sum over all the paths to obtain
accurate splittings.
The indications are that more than
one family of paths including those of order larger than the
minimal order is needed as soon as there is a significant amount
of chaos in the system. 
The situation is reminiscent of the multistep indirect paths
invoked in the annular billiard studies\cite{frido}.
This provides important clues to the underlying
mechanism of dynamical tunneling in the system. However, a semiclassical
viewpoint on these perturbative minimal paths would lead to
better insights into the process of dynamical tunneling from a phase
space viewpoint. 

It is a great pleasure to acknowledge Arul Lakshminarayan for
helpful and critical discussions.
This work was supported by funds from the Department of
Science and Technology and the Council for Scientific and Industrial
Research, India.

\end{document}